\documentclass[12pt]{article}

\begin{document}
\newcommand{\be}{\begin{equation}}
\newcommand{\ee}{\end{equation}}
\bibliographystyle{unstr}

\begin{center}

Multiplicity distributions and $H_q$ moments

\vspace{2mm}

I.M. Dremin

\vspace{1mm}

Lebedev Physical Institute, Moscow 119991, Russia

\end{center}

\begin{abstract}
It is argued that a multicomponent structure of $e^+e^-$ and $p\bar p$ high 
energy collisions determines the shapes of multiplicity distributions and
oscillations of $H_q$ moments. QCD and convoluted NBD are used for 
comparison with experimental data. Predictions at LHC and higher energies
are shown. Difference between $e^+e^-$ and $p\bar p$ is discussed. 
\end{abstract}

Multiplicity distributions are the integral characteristics of multiparticle 
production processes. They can be described either in terms of probabilities
$P(n,E)$ to create $n$ particles at energy $E$ or by the moments of 
these distributions. It has been found that their shapes possess some common 
features in all reactions studied. At comparatively low energies, these 
distributions are relatively narrow and have sub-Poissonian shapes. 
With energy increase, they widen and fit the Poisson distribution.
At even higher energies, the shapes become super-Poissonian, i.e. 
their widths are larger than for Poisson distribution. The width increases 
with energy and, moreover, some shoulder-like substructures appear. 

Their origin is ascribed to multicomponent contents of the process.
In QCD description of $e^+e^-$-processes these could be subjets formed inside 
quark and gluon jets (for the reviews see, e.g., \cite{koch}, \cite{dgar}).
In phenomenological approaches, the multiplicity distribution in a single 
subjet is sometimes approximated by the negative binomial distribution (NBD)
first proposed for hadronic reactions in \cite{giov}. For hadron-initiated 
processes, these peculiarities are also 
explained by the multicomponent structure of the process. This is either
multiladder exchange in the dual parton model \cite{cstt}, \cite{kaid}, varying number 
of clans \cite{gugo} or multiparton interactions \cite{alex}, \cite{mwal},
\cite{goul}.

Such evolution of the multiplicity distributions can be quantitatively 
described by the energy behavior of their moments. To introduce them, 
let us write the generating function of the multiplicity distribution as
\be
G(E,z) = \sum_{n=0}^{\infty }P(n,E)(1+z)^{n}.              \label{3}
\ee
In what follows, we will use the so-called unnormalized factorial ${\cal F}_q$
and cumulant ${\cal K}_q$ moments defined according to the following formulas
\be
{\cal F}_{q} = \sum_{n} P(n)n(n-1)...(n-q+1) =
 \frac {d^{q}G(E,z)}{dz^{q}}\vline _{z=0}, 
\label{4}
\ee
\be
{\cal K}_{q} = \frac {d^{q}\ln G(E,z)}{dz^{q}}\vline _{z=0}. \label{5}
\ee
They define correspondingly the total and genuine correlations among the 
particles produced (for more details see \cite{ddki}, \cite{dgar}).
Since both ${\cal F}_q$ and ${\cal K}_q$ strongly increase with
their rank and energy, the ratio
\be
H_q=\frac{{\cal K}_q}{{\cal F}_q}=1-\sum_{p=1}^{q-1}\frac{\Gamma(q)}
{\Gamma(p+1)\Gamma(q-p)}H_{q-p}\frac{{\cal F}_p{\cal F}_{q-p}}{{\cal F}_q}, 
  \label{hq}
\ee
first introduced in \cite{13}, is especially useful due to partial 
cancellation of these dependences. The factorial moments ${\cal F}_q$'s 
are always positive by definition (Eq. (\ref{4})). The cumulant moments 
${\cal K}_q$'s can change sign. They are equal to 0 for Poisson distribution.

The generating functions for quark and gluon jets satisfy definite equations
in perturbative QCD (see \cite{dkmt}, \cite{dgar}). It has been analytically 
predicted in gluodynamics \cite{13} that at asymptotically high energies 
$H_q$ moments are positive and decrease as $q^{-2}$. At $Z^0$
energy they should have a negative minimum at $q\approx 5$. Some hints to 
possible 
oscillations of $H_q$ vs $q$ at higher ranks at LEP energies were obtained 
in \cite{13}. The approximate solution of the gluodynamics equation
for the generating function \cite{41} agrees with this and predicts the 
oscillating behavior at higher ranks. These oscillations were confirmed
by experimental data for $e^+e^-$ and hadron-initiated processes first in
\cite{dabg}, later in \cite{sld} and most recently in \cite{l3}.
The same conclusions were obtained from exact solution of equations for
quark and gluon jets in the framework of fixed coupling QCD \cite{21}.
A recent exact numerical solution of the gluodynamics equation in a wide 
energy interval \cite{bfoc} coincides with the qualitative features of
multiplicity distributions described above. 

In parallel, the NBD-fits of multiplicity distributions were attempted 
\cite{gugo}, \cite{gug1}. The single NBD is parameterized as
\be
P_{NBD}(n,E)=\frac{\Gamma(n+k)}{\Gamma(n+1)\Gamma(k)}\left (\frac{m}{k}\right)^n
\left(1+\frac{m}{k}\right)^{-n-k},  \label{pnbd}
\ee
where $\Gamma $ denotes the gamma-function. This distribution has two 
adjustable parameters $m(E)$ and $k(E)$ which depend on energy.
However the simple fit by the formula (\ref{pnbd}) is valid till the 
shoulders appear. In that case, this formula is often replaced \cite{gugo}
by the hybrid NBD which simply sums up two or more expressions 
like (\ref{pnbd}). Each of them has its own parameters $m_j, k_j$.
These distributions are weighted with the energy dependent probability 
factors $w_j$ which sum up to 1. Correspondingly, the number of adjustable 
parameters drastically increases. The analytical expressions for first
five moments of two-NBD fits have been derived \cite{dnbd}.

It was proposed recently \cite{ippi} that hadron interactions can be 
represented by a set of Independent Pair Parton Interactions (IPPI-model)
and each of binary parton collisions is described by NBD with the same 
parameters. Then their convolution leads to a common distribution 
\be
P(n; m, k)=\sum_{j=1}^{j_{max}}w_jP_{NBD}(n; jm, jk).   \label{pnj}
\ee
This is the main equation of
IPPI-model. One gets a sum of negative binomial distributions with 
shifted maxima and larger widths for a larger number of collisions. 
No new adjustable parameters appear in the distribution 
for $j$ pairs of colliding partons. The probabilities $w_j$ 
are determined by collision dynamics and, in principle, can be 
evaluated if some model is adopted (e.g., see \cite{kter}, \cite{mwal}).
One can expect that at very high energies 
$w_j$ is a product of $j$ probabilities $w_1$ for one pair.
Then from the normalization condition 
\be
\sum_{j=1}^{j_{max}}w_j=\sum_{j=1}^{j_{max}}w^j_1=1   \label{w1w1}
\ee
one can find $w_1$ if $j_{max}$, which is determined by the maximum 
number of parton interactions at a given energy, is known. In asymptotics, 
$j_{max}\rightarrow \infty $ and $w_1=0.5$.

The factorial moments of the distribution (\ref{pnj}) are
\be
{\cal F}_{q} = \sum_{j=1}^{j_{max}}w_j\frac{\Gamma(jk+q)}{\Gamma(jk)}
\left(\frac{m}{k}\right)^q=\phi_q(k)\left(\frac{m}{k}\right)^q   \label{fphi}
\ee
with
\be
\phi_q(k)=\sum_{j=1}^{j_{max}}w_j\frac{\Gamma(jk+q)}{\Gamma(jk)}. \label{phiq}
\ee
The $m$-dependence of cumulant moments is similar. For $H_q$ moments one gets
\be
H_q=1-\sum_{p=1}^{q-1}\frac{\Gamma(q)}{\Gamma(p+1)\Gamma(q-p)}H_{q-p}\frac
{\phi_p\phi_{q-p}}{\phi_q}.   \label{hqph}
\ee
Thus, according to Eq. (\ref{hqph}) $H_q$ are functions of the 
parameter $k$ only and do not depend on $m$ in IPPI-model.
This remarkable property of $H_q$ moments provides an opportunity to
fit them by one parameter.

Once the parameter $k$ is found from fits of $H_q$, it is possible to get
another parameter $m$ rewriting Eq. (\ref{fphi}) as follows
\be
m=k\left(\frac {{\cal F}_{q}}{\phi_q(k)}\right)^{1/q}. \label{mkfp}
\ee
This formula is a sensitive test for the whole approach because it states
that the definite ratio of $q$-dependent functions to the power $1/q$ becomes
$q$-independent if the model is correct. Moreover, this statement should be 
valid only for those values of $k$ which are determined from $H_q$ fits.
Therefore, it can be considered as a criterion of a proper choice of $k$ 
and of the model validity, in general.
 
We have compared IPPI-model conclusions with experimental multiplicity
distributions of E735 \cite{e73} collaboration for $p\bar p$ collisions at 
energies 300, 546, 1000 and 1800 GeV extrapolated \cite{walk}, \cite{arn} to the 
full phase space. An analysis of experimental data done in \cite{walk} has shown 
that 2 parton pairs are already active at energies above 120 GeV. We assume
that 3 parton pairs are active at 300 and 546 GeV, 4 at 1000 and 1800 GeV,
5 at 14 TeV and 6 at 100 TeV with NBD for a single collision. 
We divide by 2 the charged particle multiplicity and consider multiplicity of 
particles with the same charge (say, n+).
Then, factorial and $H_q$ moments are obtained from experimental data on 
$P(n+)$ according to Eqs. (\ref{4}), (\ref{hq}).

Experimental $H_q$ moments were fitted by Eq. (\ref{hqph}) to get the 
parameters $k(E)$ of the IPPI-model. We show in Fig. 1 how perfect are 
these fits at 1.8 TeV for $k$ equal to 3.7 and 4.4. With these values of the 
parameter $k$, we have checked whether $m$ is constant as a function of $q$.
Experimental factorial moments and IPPI-values for $\phi_q$ were inserted in 
Eq. (\ref{mkfp}).
The $m(q)$ dependence is shown in Fig. 2 for the same values of $k$ and for
much larger value 7.5. The constancy of $m\approx 12.94$ is fulfilled with 
an accuracy better than 1.5$\%$ for $k=4.4$ up to $q=16$.  

The same-charge multiplicity distribution at 1.8 TeV has been fitted with 
parameters $m=12.94$ and $k$=4.4 as shown in Fig. 3. 
The similar procedure has been applied to data at energies 300, 546, 1000 GeV. 
We have found that $m$ logarithmically increases with energy.
The parameter $k$ is equal to 4.4 at 546 GeV and about 7-8 at 300 and 1000 GeV.
Extrapolating $m$ and choosing $k$=4.4 and 8, we calculated 
multiplicity distributions at 14 TeV and 100 TeV shown in Fig. 4.
Also shown, is its shape at 14 TeV if ladder model probabilities 
\cite{kter}, \cite{mwal} are inserted in (\ref{pnj}).
Oscillations of $H_q$ still persist \cite{ippi}. Their amplitudes 
($\approx 10^{-2}$) are about two orders of magnitude larger than those 
for $e^+e^-$ at $Z^0$ ($\approx 10^{-4}$). No suitable IPPI-fit by Eq. 
(\ref{hqph}) has been found for $e^+e^-$ data. The asymptotical behavior
of $H_q$ moments in IPPI-model is incompatible with $1/q^2$ dependence predicted 
by QCD. The anomalous fractal dimensions of $e^+e^-$ and $p\bar p$ also differ
\cite{ddki}.
Thus, there is no direct similarity of $e^+e^-$ and $p\bar p$ (see
\cite{ippi} for more details).

To conclude, the model of independent pair parton interactions is quite 
successful in describing experimental multiplicity distributions. Oscillations 
of $H_q$ moments can be ascribed to the mutlicomponent structure of both
lepton- and hadron-initiated processes. Predictions for LHC energies are
presented. No complete similarity of multiplicity distributions in 
$e^+e^-$ and $p\bar p$ has been observed.

This work is supported in part by the RFBR grants 02-02-16779,
03-02-16134, NSH-1936.2003.2.\\

Figure captions.\\

Fig. 1 A comparison of $H_q$ moments derived from experimental data at 1.8 TeV 
(squares) with their values calculated with parameter $k$=4.4 
(dash-dotted line) and 3.7 (solid line).

Fig. 2 The $q$-dependence of $m$ for $k$=4.4 (squares), 3.7 (circles) and 7.5 (triangles).

Fig. 3 The same-charge multiplicity distribution at 1.8 TeV, its fit at $m=12.94$,
       $k=4.4$ (solid line). The dash-dotted line demonstrates what 
       would happen if NBD is replaced by Poisson distribution.

Fig. 4 The same-charge multiplicity distributions at 14 TeV and 100 TeV predicted
       by IPPI-model with logarithmic extrapolation of $m$ (solid line - 14 TeV, 
       $k$=4.4; dash-dotted - 14 TeV, $k$=8; dashed - 100 TeV, $k$=4.4; 
       for the ladder model: dotted - 14 TeV, $k$=4.4).\\


\begin{thebibliography}{99}
\bibitem{koch}
V.A. Khoze, W. Ochs, Int. J. Mod. Phys. A  12 (1997) 2949.
\bibitem{dgar}
I.M. Dremin, J.W. Gary, Phys. Rep. 349 (2001) 301.
\bibitem{giov}
A. Giovannini, Nuovo Cim. A 15 (1973) 543. 
\bibitem{cstt}
A. Capella, U. Sukhatme, C.I. Tan, J. Tran Thanh Van, Phys. Lett. B 81 
(1979) 68.
\bibitem{kaid}
A.B. Kaidalov, Phys. Lett. B 116 (1982) 459.
\bibitem{gugo}
A. Giovannini, R. Ugoccioni, Phys. Rev. D 59 (1999) 094020.
\bibitem{alex}
E735 Collaboration, T. Alexopoulos et al., Phys. Lett. B 435 (1998) 453.
\bibitem{mwal}
S.G. Matinyan, W.D. Walker, Phys. Rev. D 59 (1999) 034022.
\bibitem{goul}
K. Goulianos, Phys. Lett. B 193 (1987) 151.
\bibitem{ddki}
E.A. De Wolf, I.M. Dremin, W. Kittel, Phys. Rep. 270 (1996) 1.
\bibitem{13}
I.M. Dremin, Phys. Lett. B 313 (1993) 209.   
\bibitem{dkmt}
Yu.L. Dokshitzer, V.A. Khoze, A.H. Mueller, S.I. Troyan, {\it Basics of
perturbative QCD} ed. by J. Tran Thanh Van
(Gif-sur-Yvette, Editions Frontieres, 1991).
\bibitem{41}
I.M. Dremin, V.A. Nechitailo, Mod. Phys. Lett. A  9 (1994) 1471;
JETP Lett. 58 (1993) 881.
\bibitem{dabg}
I.M. Dremin, V. Arena, G. Boca et al., Phys. Lett. B 336 (1994) 119.
\bibitem{sld}
SLD Collaboration, K. Abe et al., Phys. Lett. B 371 (1996) 149.
\bibitem{l3}
L3 Collaboration, P. Achard et al., Phys. Lett. B 577 (2003) 109.
\bibitem{21}
I.M. Dremin, R.C. Hwa, Phys. Rev. D 49 (1994) 5805; Phys. Lett. 
B 324 (1994) 477.
\bibitem{bfoc}
M.A. Buican, C. F\"{o}rster, W. Ochs, Eur. Phys. J. C 31 (2003) 57.
\bibitem{gug1}
A. Giovannini, R. Ugoccioni, hep-ph/0312205.
\bibitem{dnbd}
I.M. Dremin, hep-ph/0401013.
\bibitem{ippi}
I.M. Dremin, V.A. Nechitailo, hep-ph/0402286.
\bibitem{kter}
A.B. Kaidalov, K.A. Ter-Martirosyan, Phys. Lett. B 117 (1982) 247.
\bibitem{e73}
E735 Collaboration, F. Turkot et al., Nucl. Phys. A 525 (1991) 165.
\bibitem{walk}
W.D. Walker, Phys. Rev. D 69 (2004) 034007.
\bibitem{arn}
UA5 Collaboration, G. Arnison et al., Phys. Lett. B 118 (1982) 167.

\end{thebibliography}
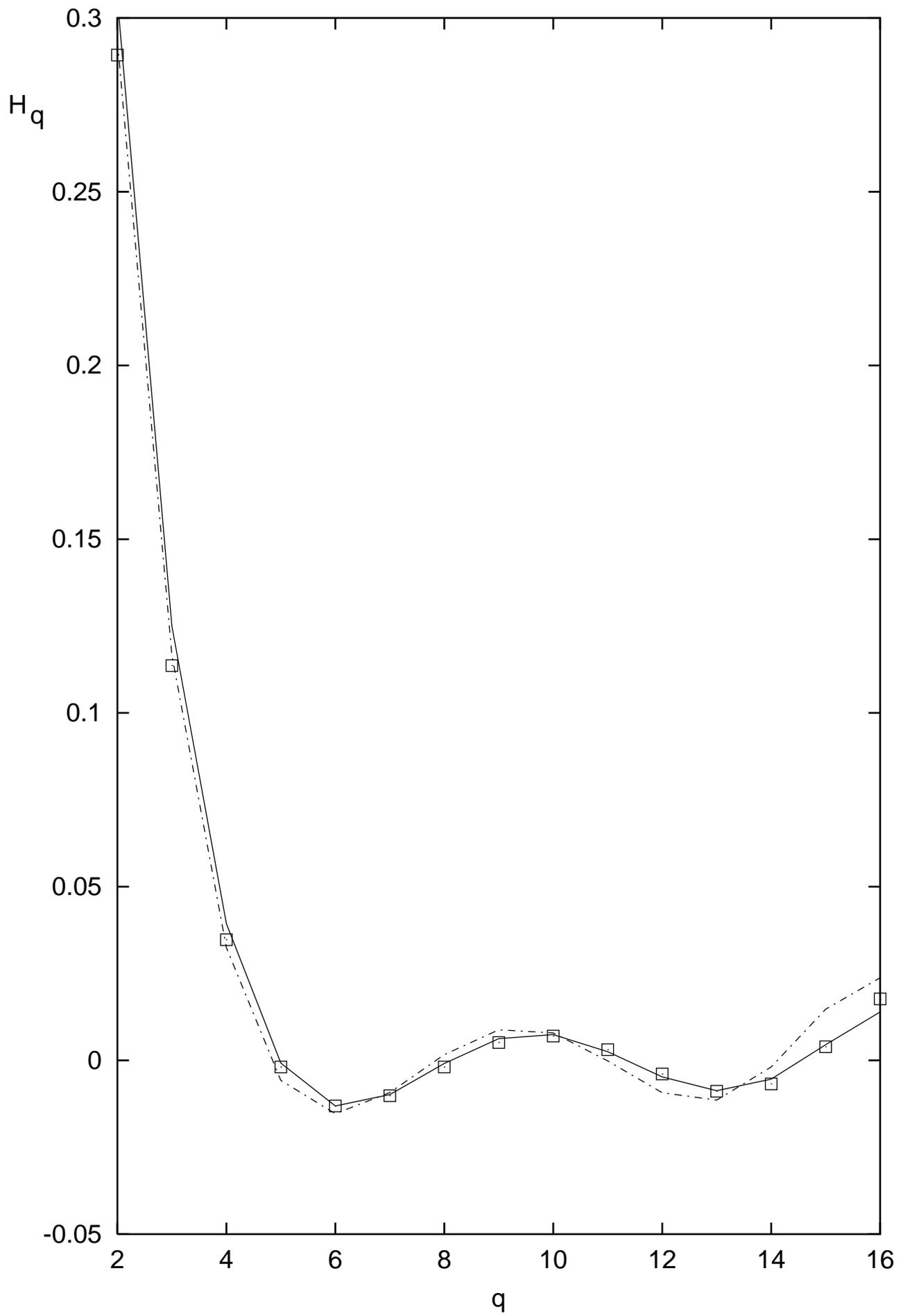
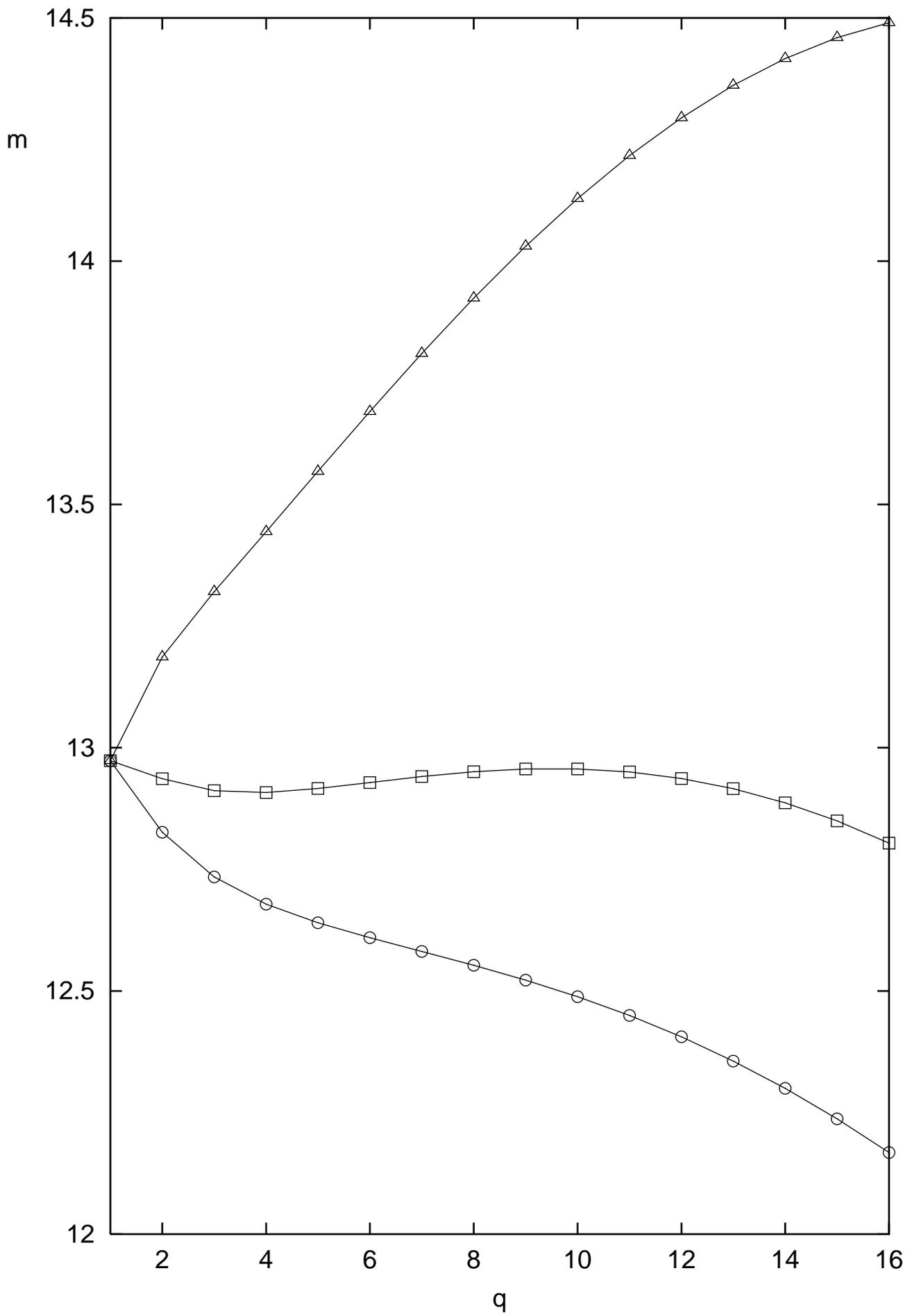
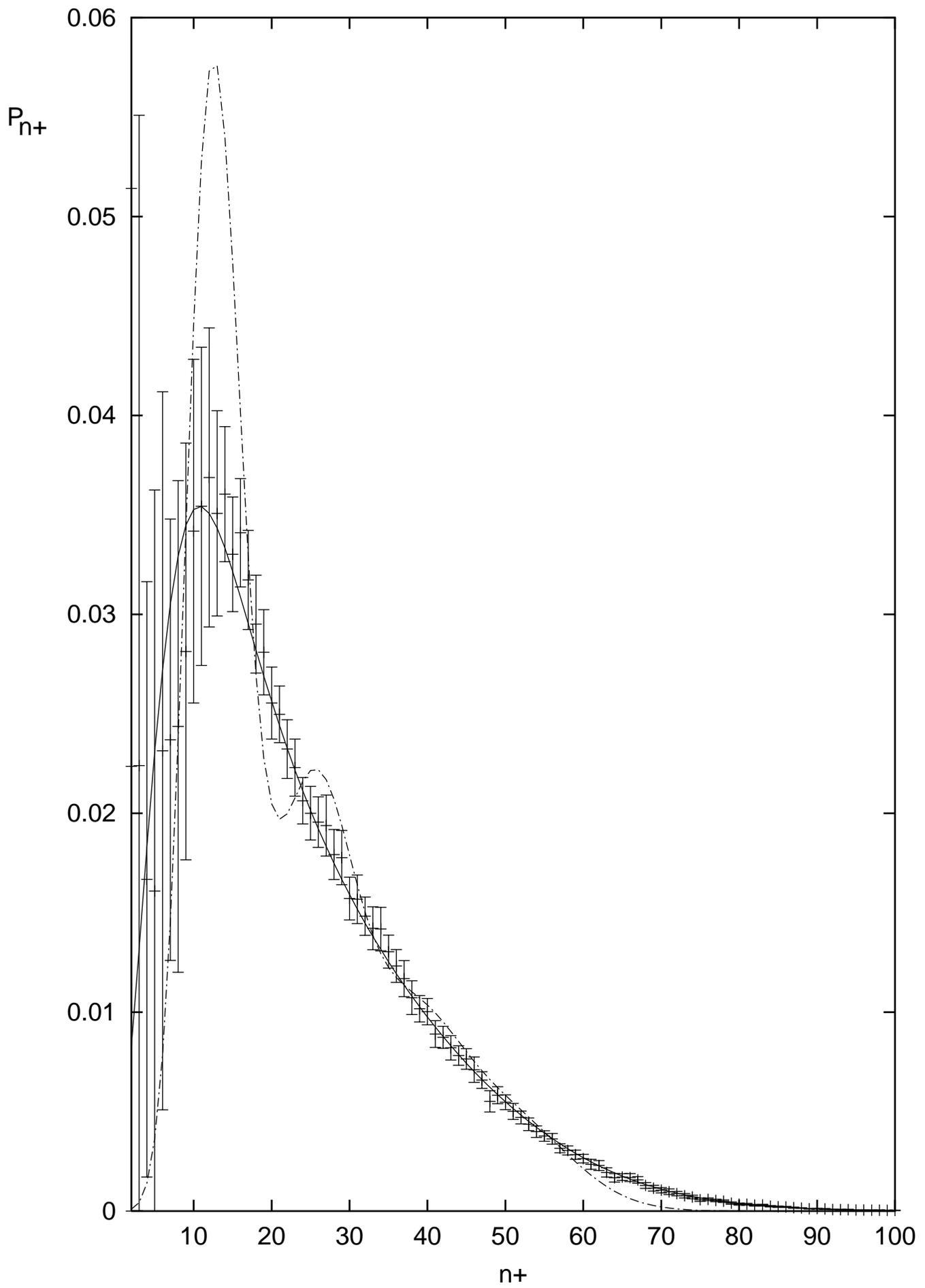
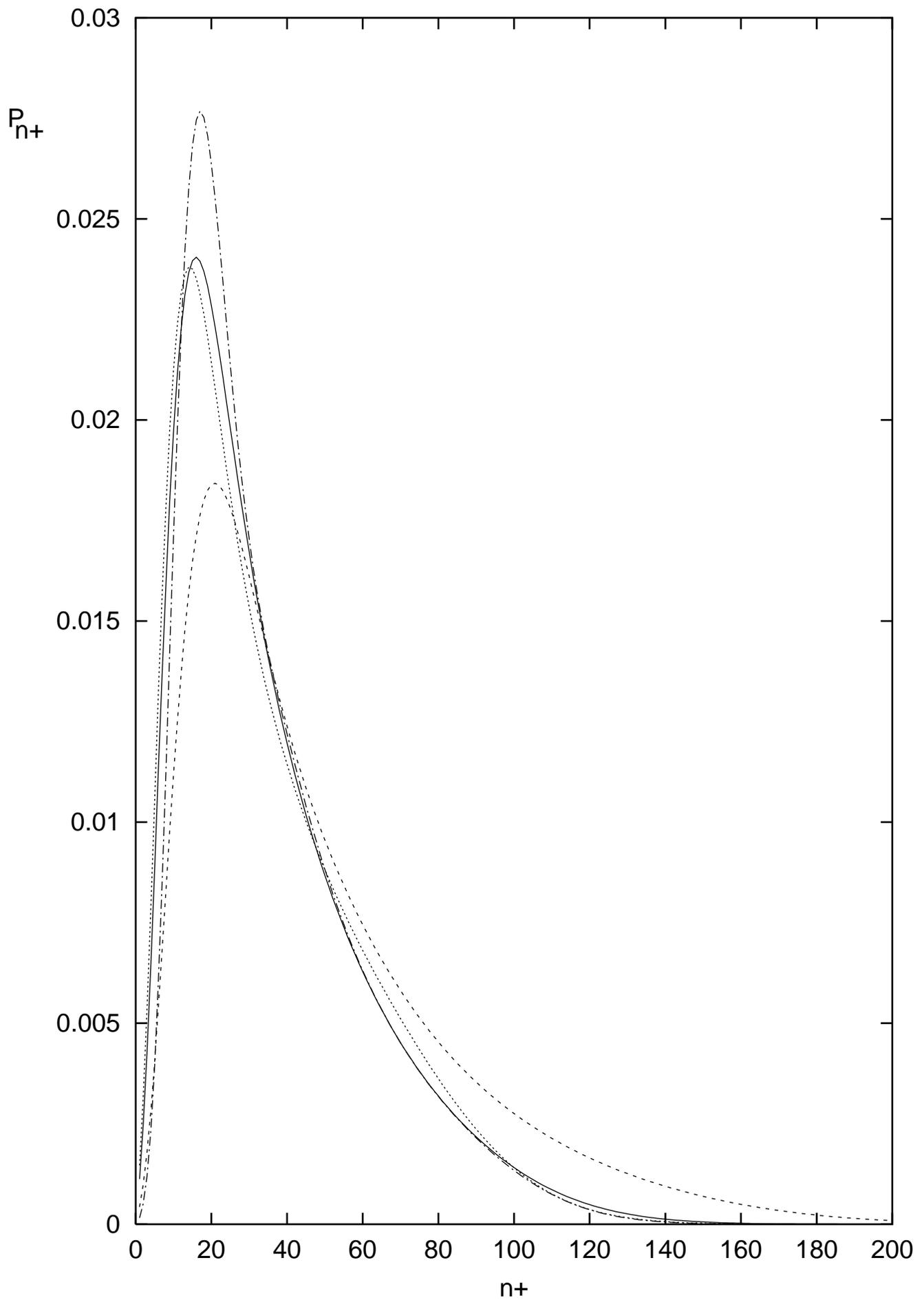
\end{document}